\documentclass[12pt, twoside, here]{article}
\usepackage{epsf}
\usepackage{times,colordvi,amsmath,epsfig,float,color,multicol}
\usepackage{graphics}
\usepackage{hhline}
\usepackage[large]{subfigure}
\usepackage[latin1]{inputenc}
\usepackage{rotating}
\usepackage{marvosym}
\usepackage{amsfonts}
\usepackage{amssymb}

\title{A Set of Discrete Formulae for the Performance of a Tsetse Population During Aerial Spraying}

\renewcommand{\thefootnote}{\fnsymbol{footnote}}

\author{S. J. Childs \\ \\ {\small\em Department of Mathematics and Applied Mathematics, University of the Free State,} \\ {\small\em P.O. Box 339, Bloemfontein, 9300, South Africa.} \\ {\small\em Tel: +27 51 4013386 \ \ Email: simonjohnchilds@gmail.com}}

\date{Acta Tropica, 125: 202--213, 2013}

\begin{document}

\maketitle
\renewcommand{\thefootnote}{\arabic{footnote}}

\begin{abstract}
\noindent {\em A set of discrete formulae that calculates the hypothetical
impact of aerial spraying on a tsetse population is derived and the work is
thought to be novel. Both the original population and the subsequent generations
which survive the aerial spraying, may ultimately be thought of as deriving from
two, distinct sources. These origins are, however, neither distinct, nor
relevant by the third generation. It is for this reason that the female
population is considered to be composed of the following four categories for the
purposes of derivation: Original flies which existed as such at the commencement
of spraying; original pupae which existed as such at the commencement of
spraying; the immediate descendants of both the aforementioned categories,
during spraying; third and higher generation descendants. In theory, the latter
category is a recurrence relation. In practice, the third generation's pupal
stage has hardly come into existence, even by the end of a completed operation.
Implicit in the formulae is the assumption of one, temperature-dependent
mortality rate for the entire pupal stage, a second for the period between
eclosion and ovulation and yet a third for the entire, adult life-span. Gravid
female resistance to the insecticide is assumed to be inconsequential. A further
assumption of the formulae is that at least one male is always available (degree
of sterility variable).} 
\end{abstract}


\section{Introduction}

The {\em Glossina} genus is the vector of trypanasomiasis in Africa. There are
about thirty three species and subspecies of tsetse fly, whereas about half as
many trypanosomes of the salivarian clade are thought to exist (Gooding and
Krafsur\nocite{GoodingAndKrafsur}, 2004; Stevens and
Brisse\nocite{StevensAndBrisse}, 2004). Thirty six African countries are still
afflicted by human, African trypanosomiasis (HAT), although nagana is still
of veterinary and economic importance in others e.g. South Africa
(Anonymous\nocite{WHO1}, 2012). 

The most common causes of nagana in livestock are {\em Trypanosoma congolense}
and {\em T. vivax}, in that order of priority. Neither pathogen has ever been
known to infect an human host, although both domestic and wild animals serve as
the reservoir for the human afflictions, {\em T. brucei gambiense} and {\em T.
brucei  rhodesiense}. {\em T. gambiense} is associated with chronic disease in
West Africa, while {\em T. rhodesiense} is associated with acute disease in East
Africa. Although the advance of HAT is spectacularly rapid in the case of
\mbox{\em T. rhodesiense}, \mbox{\em T. gambiense} can be dangerously insidious,
the symptoms often only becoming manifest once it is too late to treat (95\% of
all HAT cases are attributed to \mbox{\em T. gambiense} according to
Anonymous\nocite{WHO1}, 2012). Not enough is known about the vector competence
of the various tsetse species, as was recently illustrated by the findings of
Motloang et al.\nocite{MotloangMasumuVanDenBosscheMajiwaLatif} (2009). The {\em
fusca} and {\em palpalis} groups are largely confined to West Africa, while the
{\em morsitans} group is largely confined to the Eastern side of the continent,
a few exceptions to this rule occurring in both the {\em fusca} and {\em
morsitans} groups. Three members of the {\em morsitans} group, namely {\em
Glossina morsitans}, {\em Glossina pallidipes} and {\em Glossina austeni} could
be considered to be mainly problematic in the Southern and East African
theatres, while the problem assumes a far greater diversity around Lake Victoria
and to the west of it. Members of the {\em palpalis} group are notorious vectors
in the West African theatre ({\em Glossina fuscipes fuscipes}, in particular,
having been implicated by the focus of numerous epidemics).

Trypanosomiasis is regarded by many African countries to be largely of
veterinary and economic importance, in modern times. This has certainly not
always been the case and Leak\nocite{Leak1} (1999) provides a grim reminder that
in the opening years of the twentieth century, around \mbox{200 000} people died
of trypanosomiasis in the provinces of Buganda and Busoga alone and that these
provinces eventually had to be evacuated. F\`{e}vre et
al.\nocite{FevreColemanWellburnMaudlin} (2004) put the figure closer to around
two thirds of the lake-shore population, for a slightly longer period and the
epidemic reached similar proportions in the Congo river basin
(Anonymous\nocite{WHO1}, 2012). Today HAT has all but vanished, largely as a
result of the all-out war waged against tsetse during the twentieth century. So
great has been the success that, in 2010, only 7139 new cases were reported, the
biggest contributor being the Democratic Republic of the Congo
(Anonymous\nocite{WHO1}, 2012). 

This is no small achievement and such success has not come without a price.
\mbox{Du Toit}\nocite{DuToit} (1954) put the cost of {\em G. pallidipes}
eradication from KwaZulu-Natal, in the first half of the twentieth century, at
well in excess of \mbox{\pounds 100 000}. Properly planned aerial spraying has
proved to be the most effective means of tsetse control and it is with the
prohibitive costs in mind that this research attempts to make operations as
efficient and successful as possible. The modern operation conventionally
utilizes a relatively harmless pyrethroid such as endosulfan or deltamethrin
(Allsopp\nocite{Allsopp1}, 1984). An aerosol of insecticide is discharged from a
formation of aircraft, flying at low altitude (less than 100 metres a.g.l.) and
guided by G.P.S. Adult flies are extremely susceptible to the insecticide and
kill rates very close to 100\% can be anticipated under favourable conditions. 

The main challenge to controlling tsetse by aerial spraying is that the pupal
stage is largely protected from insecticides. Repeat spray cycles therefore need
to be scheduled to kill new flies which begin emerging immediately after
spraying. Both economic and environmental considerations dictate that the number
of such cycles be minimised. The problem, however, is that if the time between
spray cycles is too long, recently eclosed flies will themselves mature, become
reproductive and larviposit. The underlying philosophy to the aerial spraying of
tsetse relies heavily on the fact that all developmental periods in the {\em
Glossina} genus are entirely temperature dependent and are therefore readily
predictable. By knowing the mean temperature spray cycles can be scheduled two
days short of the time to the production of the first larva; the two-day safety
margin being designed to ensure that there is no variance in
time-to-first-larva to levels below the length of the spray cycle.

The strategy explored in Childs\nocite{Childs4} (2011) was one in which the
repeated spray cycles are continued until two sprays subsequent to the eclosion
of the last, pre-spray-larviposited, female pupae. None of the observations in
that work are, however, valid in the event that the operation is terminated one,
or more, sprays short and there could be many reasons for pursuing such a
strategy in the modern scenario. Costs, environmental considerations and an
area-wide, integrated approach to pest management which contemplates the use of
the sterile insect technique (Barclay and Vreysen\nocite{BarclayAndVreysen},
2010), are only a few of the reasons why a curtailed operation has increasingly
been entertained as a `knock-down', rather than as an agent of eradication, in
recent times. A recent shift in interest from savannah, to the more
inaccessible, riverine and forested habitats, in combination with a better
understanding of odour-baited targets, pour-ons and dips
(Childs\nocite{Childs3}, 2010 and Esterhuizen et
al.\nocite{EsterhuizenKappmeierGreenNevillVanDenBossche} 2006), has led to these
alternative counter measures recently having been assigned a far more
significant role in control and eradication, than in the past. In the event that
spray cycles are not continued for the full duration, the formula derived in
Childs\nocite{Childs4} (2011) is not appropriate. A more comprehensive set of
formulae is required, one which, for example, also accounts for other categories
of pupae, such as the immediate, pupal descendents of pre-spray-existing
flies, as well as actual flies themselves. 

The effect of temperature on aerial spraying, through the reproductive cycle and
general population dynamics of the tsetse fly, can easily be taken into account.
The same cannot, however, be said for the effect of temperature on spray
efficacy, it being a property unique to each and every environment and the
conditions prevailing at the time. Very high kill rates usually (though not
always) come about as a result of the sinking air associated with cooler
weather. It favours the settling of insecticidal droplets. The inherent toxicity
of deltamethrin and many other pyrethroids also decreases with temperature,
contrary to the toxicity of most insecticides. The effects of anabatic winds,
the protection afforded by the forest canopy and multifarious other variables,
are just as relevant to spray efficacy. No account is taken of the mechanism in
gravid females, whereby lipophilic toxins are excreted, sacrificing larvae in
utero for survival, either. The effect of temperature and age on spray efficacy
is therefore not modelled and it is, instead, a variable in the formulation.
Spray efficacy is usually measured in the field, with hindsight, rather than
predicted. Three kill rates of around 99\%, 99.9\% and 99.99\% respectively are
entertained in this work. They should be thought of as being broadly associated
with the warmer, intermediate and cooler parts of the low-temperature range
respectively. It is in this way that the hypothetical impact of aerial spraying
on tsetse fly populations is formulated.

The formulae derived in this work are largely a predictive tool. They provide a
convenient means of calculating theoretical levels of control in the aerial
spraying of tsetse, by way of spreadsheets and simple algorithms, in which the
outcome is based on mean temperature and spray efficacy. They also provide a
convenient means of making `back-of-an-envelope' estimates based on first
order terms. The formulae provide a means to calculating the outcome at mean
temperature. The data presented in Hargrove\nocite{Hargrove1} (1990), for
example, suggest that the temperature in tsetse environments often varies
little. The restriction to mean temperature is not problematic from a point of
view of prediction, since one can usually only forecast mean temperature. An
algorithm is the next logical step, brought about by the introduction of
variable temperature.

\section{Aerial Spraying and the Life-Cycle of the Tsetse Fly} \label{lifeCycle}

The female tsetse fly mates only once in her life with the chance $\eta$ that
she is successfully inseminated ($\eta$ is usually taken to be unity). She also
produces only one larva at a time. The time between female eclosion and the
production of the first larva is known as time-to-first-larva, $\tau_1$.
Thereafter she produces pupae at a shorter interlarval period, $\tau_2$. The
effect of temperature on the first and subsequent interlarval periods has been
estimated in the field, using {\em G. pallidipes}. The predicted mean time taken
from female eclosion to the production of the first pupa is obtained using
Jackson's (Anonymous\nocite{Anonymous}, 1955) temperature-dependent formula,
\begin{eqnarray} 
\tau_i &=& \frac{ 1 }{ k_1 + k_2 \left( T - 24 \right) } \hspace{10mm} i = 1,2, \nonumber
\end{eqnarray} 
in which $k_1 = 0.061$ and $k_2 = 0.0020$
(Hargrove\nocite{Hargrove4}\nocite{Hargrove5}, 1994 and 1995). The subsequent
interlarval periods are predicted using $k_1 = 0.1046$ and $k_2 = 0.0052$
(Hargrove\nocite{Hargrove4}\nocite{Hargrove5}, 1994 and 1995). The interlarval
periods are therefore entirely temperature dependent and readily predictable.
Use of this formula needs, however, to be tempered by a knowledge of the large
standard deviation presented in Hargrove (1994\nocite{Hargrove4} and
1995\nocite{Hargrove5}), as well as the fact that larviposition usually takes
place in the late afternoon, for {\em G. morsitans} (Potts\nocite{Potts1}, 1933,
reported by Jackson\nocite{Jackson1}, 1949, and Brady\nocite{Brady1}, 1972), or
afternoon shade in the case of {\em G. palpalis} (Jackson\nocite{Jackson1},
1949, and Buxton\nocite{Buxton1}, 1955). There exists an ever present risk in
interpreting the output to have a precision any better than the daily cohort and
a discrete model may be more appropriate than a continuous one under these
circumstances. 

What is the relevance of the above formula? Since the pupae present in the
ground are unaffected by insecticide, the idea is to schedule follow-up
operations shortly before the first flies to eclode, after spraying, themselves
mature and become reproductive. Subsequent sprays are consequently scheduled two
days short of the time to the first larva. This length of the spray interval is
denoted $\sigma$ in the formulae to follow. For temperatures of $22 \
^\circ\mathrm{C}$ and below, both Jackson's curve and the data reported in
Hargrove\nocite{Hargrove3} (2004) suggest that spraying two days before the time
to first larva (the one predicted using the Hargrove, 1994\nocite{Hargrove4} and
1995\nocite{Hargrove5}, coefficients) is sufficient to ensure that none of the
recently eclosed female flies ever give birth prior to being sprayed. This
observation is supported by the success of operations such as those of Kgori et.
al\nocite{Torr1} (2006). Caution may, however, need to be exercised in the case
of {\em G. austeni}, in that both periods could be shorter than the above
formula predicts. This suspicion is based on the small size of the fly and in
keeping with its shorter puparial duration (Parker\nocite{Parker1}, 2008). A
shorter time between eclosion and the production of the first larva is a concern
for the aerial spraying of {\em G. austeni}. For {\em G. brevipalpis} one
suspects longer periods based on diammetrically opposite arguments. The only
relevance to aerial spraying in this latter case is economic, inefficiency being
the only expected consequence. 

Pupae that are successfully larviposited remain in the ground for a period of
time. The duration of the period between larviposition and the emergence of the
first imago is known as the puparial duration and is denoted $\tau_0$ in this
work. The puparial duration is also a function of temperature and may be
predicted using the formula
\begin{eqnarray} \label{1}
\tau_0 &=& \frac{ 1 + e^{a + bT} }{k} \nonumber
\end{eqnarray} 
(Phelps and Burrows\nocite{phelpsAndBurrows1}, 1969). For females, $k = 0.057
\pm 0.001$, $a = 5.5 \pm 0.2$ and $b = -0.25 \pm 0.01$
(Hargrove\nocite{Hargrove3}, 2004). The fact that pupae usually eclode in the
evening (Vale et. al.\nocite{ValeHargroveJordanLangleyAndMews} 1976) again begs
the question of over-interpreting precision. There exists an ever present risk
in interpreting the output to have a precision greater than the daily cohort and
a discrete model is again indicated as being more appropriate than a continuous
one. 

What is the relevance of the above formula? A cautious strategy advocates that
spray cycles should be repeated until after the last pre-spray-larviposited
pupae eclode and it is safer to continue until at least two sprays after their
eclosion due to variation in the environment. If, under such circumstances, $s$
denotes the total number of sprays, the total duration of the entire spraying
operation is $s - 1$ cycles. Again, caution needs to be exercised in that
Parker\nocite{Parker1} (2008) reports that {\em G. brevipalpis} takes a little
longer than the above formula predicts, whereas the puparial durations of all
other species are thought to lie within 10\% of the value predicted. For the
same conditions which produce a {\em G. morsitans} puparial duration of 30 days,
{\em G. brevipalpis} has a puparial duration of 35 days. This has important
implications for the aerial spraying of {\em G. brevipalpis}. The shortest
puparial duration is that of {\em G. austeni}. {\em G. austeni}'s puparial
duration was 28 days under the aforementioned conditions. These observations are
noteworthy given the South African context of a sympatric, {\em G.
brevipalpis}-{\em G. austeni} population.

Other aspects of tsetse population dynamics are also largely
temperature-dependent \\ (\mbox{Hargrove\nocite{Hargrove3}, 2004}), although
soil-humidity can play an as, or more, important role in early mortality,
depending on the species (Childs\nocite{Childs2}, 2009). While the effects of
both temperature and humidity on pupal mortality are known to be important, they
vary profoundly according to the exact stage of development and are cumulative,
rather than instantaneous. One might therefore surmise that the age-dependence
which characterises post-pupal mortality (observed by
Hargrove\nocite{Hargrove1}, 1990 and 1993\nocite{Hargrove2}) is largely a
consequence of pupal history. Fortunately, variables such as soil-humidity and
vegetation index have little to do with metabolic rate, hence the timing of
spray cycles, and worst-case values might therefore be used. Alternatively,
they can be regarded to vary (and therefore be relevant) only in the medium to
long term. In many regions, the level of humidity and temperature are sometimes
linked. Pupae are therefore taken to die off at some temperature-dependent,
daily rate, $\delta_0$, and those flies which subsequently emerge have a
probability $\gamma$ of being female. Some comfort can be taken from the
knowledge that the effects of natural mortalities are very small in comparison
to those due to aerial spraying. They have little bearing on the overriding
trends and, to a certain extent, this knowledge permits a primitive approach.
The question of pupal mortality can also be substantially avoided through the
use of a steady-state eclosion rate, $\beta$. Hargrove\nocite{Hargrove3} (2004)
suggests adult mortality to be predictable, almost entirely
temperature-dependent, and a knowledge of post-eclosion mortalities infers the
eclosion rate and vice versa, assuming the population to be in equilibrium. It
is with this wisdom in mind that the derivation will commence. 

During the first few hours subsequent to eclosion, the young, teneral fly's
exoskeleton is soft and pliable, its fluid and fat reserves are at their lowest
and a first blood meal is imperative for its survival. It is at this time that
the insect is at its most vulnerable and it is also at this time that its
behaviour is least risk averse (Vale\nocite{Vale3}, 1974). Post-pupal survival
can be defined as $e^{- \delta_1}$ per day for the period between female
eclosion and ovulation. Thereafter the female tsetse fly's chances of survival
are higher and can be defined as $e^{ - \delta_2}$ per day.

The accumulated mortality described above can be modelled linearly as 
\begin{eqnarray} \label{2} 
\delta(t, T) & = & \left\{ 
\begin{array}{l} \delta_0 \ t  \\ 
\delta_1 \left(t  - \tau_0\right) + \delta_0 \tau_0  \\ 
\delta_2 \left[ t - (\tau_1 - \tau_2) - \tau_0 \right] + \delta_1 \left( \tau_1 - \tau_2 \right) + \delta_0 \tau_0 
\end{array} \hspace{1mm} \mbox{for} \hspace{1mm} 
\begin{array}{rcl}
&t&< \ \tau_0 \\
\tau_0 \ \le&t&< \ \tau_1 - \tau_2 + \tau_0 \\ 
&t&\ge \ \tau_1 - \tau_2 + \tau_0, 
\end{array} 
\right. \nonumber
\end{eqnarray} 
where $t$ denotes age, for the present. For the purposes of later brevity, it is convenient to define a second cumulative mortality, one which commences at eclosion. If $t$ denotes the time elapsed since eclosion, then
\begin{eqnarray}
\delta^*(t, T) & = & \left\{ 
\begin{array}{l}
\delta_1 t \\ 
\delta_2 \left[ t - (\tau_1 - \tau_2) \right] + \delta_1 \left( \tau_1 - \tau_2 \right) 
\end{array} \hspace{5mm} \mbox{for} \hspace{5mm} 
\begin{array}{rcl}
&t&< \ \tau_1 - \tau_2 \\ 
&t&\ge \ \tau_1 - \tau_2,
\end{array} 
\right. \nonumber
\end{eqnarray}
is the aforementioned mortality desired. Some actual values of the various
mortalitites, their associated temperatures and the justification for their
selection can be found in Childs\nocite{Childs4} (2011). 

The spray-survival rate will, in contrast, be assumed to be independent of age,
whereas, in actual fact, a mechanism in gravid females exists whereby lipophilic
toxins are excreted, sacrificing larvae in utero for the mother fly's own
survival. The older the fly, the more developed this mechanism is usually found
to be. The dependence of spray efficacy on age has been ignored for two reasons.
Firstly, one might reason that a simple trade-off exists between a fly living
and a larva dying and further pregnancies should similarly terminate in
spontaneous abortion. Secondly, the spray-survival rate is a small number.
Whatever the exact value of $\phi$ may be for these older flies, the value of
$\phi^s$, or similar, should ensure that such cohorts are decimated by the end
of the operation. Of course, ignoring gravid female resistance to the
insecticide may result in slightly altered eclosion rates and the use of
inappropriate natural mortalities. Some comfort can be taken from the knowledge
that the effects of natural mortalities are very small in comparison to those
due to aerial spraying. They have little bearing on the outcome. 

\section{Strategy for Derivation}

The emphasis in the derivation is on the female population, since the male
tsetse fly's role in reproduction is relatively insignificant. At least one male
is always assumed to be available and any level of sterilization is accounted
for by way of a probability of insemination. 

A strong case obviously exists for taking the time of the first spray to be
zero, rounding the outputs of the aforementioned formulae to the nearest integer
cohort and, consequently, developing a discrete model, one in which spraying
occurs subsequent to both larviposition and eclosion on days when spraying is
relevant. Just some of the factors which recommend such an approach are that
eclosion occurs in the afternoon or early evening when the challenges of
dehydration are lower, that larviposition usually occurs not very long before
that, or coincides with it, that aerial spraying is best carried out at night
when low temperatures favour the settling of insecticidal droplets, the daily
character of most traditionally available data and unexplained variance.

\begin{figure}[H]
   \begin{center}
\includegraphics[height=7cm, angle=0, clip = true]{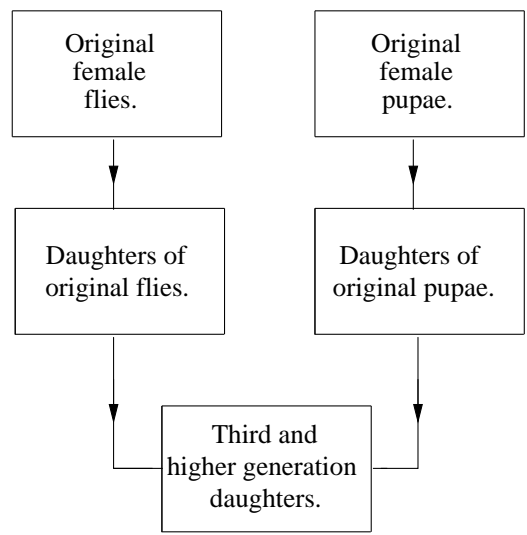}
\caption{The tsetse population is deconstructed into logically natural categories for the purposes of formulation.} \label{decomposition}
   \end{center}
\end{figure}
Both the original population and the subsequent generations which survive the
spraying, may be thought of as ultimately deriving from two distinct sources
(refer to Fig. \ref{decomposition}). These origins are, however, neither
distinct, nor relevant by the third generation. To understand why this is so,
cognizance should be taken of the fact that the number of flies in a given
cohort depends on the number of mothers which survived long enough to
successfully larviposit, not just on the cohort's own chances of survival since
larviposition. There are two distinctly different ancestral origins for the
second generation, since mothers existed as either a pupa, or a fly at the
commencement of spraying. The same is not true for third generation cohorts
since the only generalisation that can be made is that all mothers simply
eclosed sometime, subsequent to one puparial duration into the operation, and
happened to larviposit on the same day. 
\begin{table}[H]
\begin{center}
\begin{tabular}{c|c|l}  
&  &  \\
symbol \ & \ unit \ & \hspace{30mm} description \\ & & \\ \hline & & \\
N \ & \Female & \ original, steady-state, equilibrium number of females \ \\ & & \\
$\eta$ & \ - \ & \ probability of insemination \ \\ & & \\
$\beta$ & \ flies \ \Female $^{-1}$ $\mathrm{day}^{-1}$ \ & \ eclosion rate \\ & & \\
$\gamma$ & \ \Female \ $\mbox{flies}^{-1}$ \ & \ sex ratio \ \\ & & \\
$\delta_0$ & \ $\mathrm{day}^{-1}$ \ & \ puparial mortality \ \\ & & \\
$\delta_1$ & \ $\mathrm{day}^{-1}$ \ & \ post-puparial, pre-ovulatory mortality \ \\ & & \\
$\delta_2$ & \ $\mathrm{day}^{-1}$ \ & \ adult mortality \ \\ & & \\
$\tau_0$ & \ $\mathrm{days}$ \ & \ puparial duration \ \\ & & \\
$\tau_1$ & \ $\mathrm{days}$ \ & \ time between eclosion and first larva \ \\ & & \\
$\tau_2$ & \ $\mathrm{days}$ \ & \ interlarval period \ \\ & & \\
$\sigma$ & \ $\mathrm{days}$ \ & \ length of a spray cycle \ \\ & & \\
$s$ & \ $\mathrm{sprays}$ \ & \ total number of sprays \ \\ & & \\
$\phi$ & \ - \ & \ probability of surviving a single spray \ \\ & & \\
$\breve t$ & \ $\mathrm{days}$ \ & \ time to eclosion since first spray \ \\ & & \\
$E_{\mbox{\scriptsize pre-spray}}(\breve{t})$ & \ $\mathrm{flies}$ \ & \
time-$\breve{t}$-ecloding cohort which existed as \\ & & \ pupae at the commencement of spraying \ \\ & & \\
$E_a(\breve{t})$ & \ $\mathrm{flies}$ \ & \ time-$\breve{t}$-ecloding cohort,
larviposited by original, \\ & & \ adult females during spraying (second generation) \ \\ & &
\\
$E_{ps}(\breve{t})$ & \ $\mathrm{flies}$ \ & \ time-$\breve{t}$-ecloding
cohort, larviposited by original, female \\ & & \ pupae which existed as such at the commencement of \\ & & \ spraying (second generation) \\ & & \\
$E_{is}(\breve{t})$ & \ $\mathrm{flies}$ \ & \ time-$\breve{t}$-ecloding
cohort, immediately descended from \\ & & \ inter-spray-deposited, female
pupae (third generation \\ & & \ and higher) \\ & & \\
\end{tabular}
\caption{Symbols used and the quantities they denote.} \label{symbols}
\end{center}
\end{table}
\begin{figure}[H]
    \begin{center}
\includegraphics[height=19cm, angle=0, clip = true]{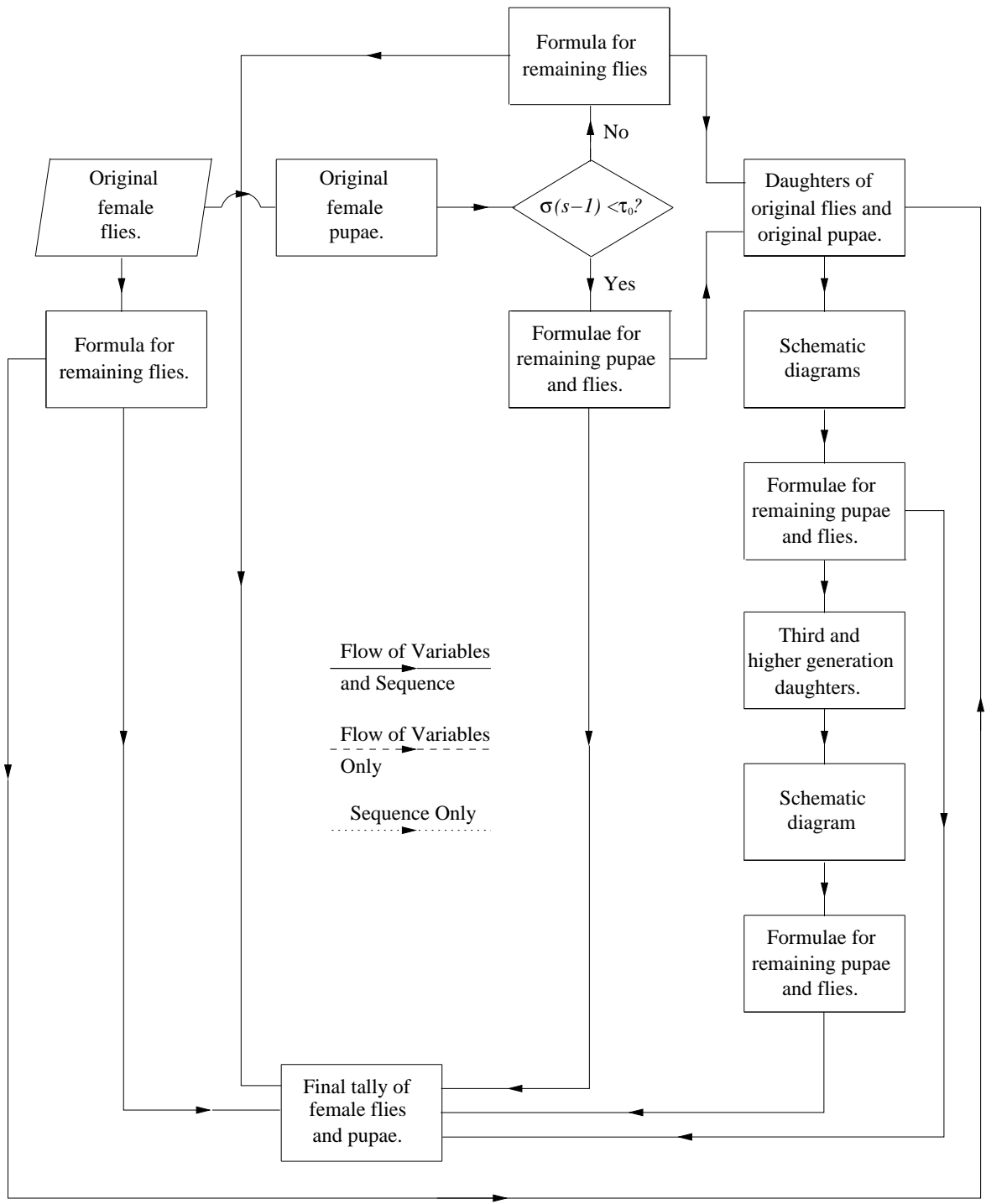}
\caption{Flow chart of both the strategy for obtaining the complete set of formulae and the calculation itself.} \label{strategy}
   \end{center}
\end{figure}
The observation of two initially distinct origins is, in some sense, an artefact
of having seperate, `start-up' pupal and fly populations, something which was
rendered possible by the assumption of an equilibrium. Yet the existance of that
equilibrium prior to spraying is no artefact. 

For the reason that there are originally two distinct sources, it is expedient
to deconstruct the surviving population into the following, categories, in the
derivation:
\begin{enumerate}
\item\label{survivingAdults} Original, female flies which existed as such at the commencement of spraying.
\item\label{preSprayExistingPupae} Original, female pupae which existed as such at the commencement of spraying.
\item\label{interSprayPupae} Daughters larviposited after the commencement of spraying, including: 
\begin{enumerate}
\item Daughters of \ref{survivingAdults} above.
\item\label{daughtersOfPreSprayExistingPupae} Daughters of
\ref{preSprayExistingPupae} above.
\item Third generation and higher daughters of this self-same category, \ref{interSprayPupae} above. 
\end{enumerate}
\end{enumerate}
Fig. \ref{strategy} summarises the strategy for both formulation and calculation.

\section{Surviving Flies}

The actual flies themselves, as distinct from pupae, which survive the last
spray are usually of no real consequence to the outcome of spraying
(Childs\nocite{Childs4}, 2011). This is not necessarily the case in instances in
which the operation has been curtailed, or kill rates are low. The state of the
adult fly population during spraying is, nonetheless, what ultimately determines
the size of the remnant population at the end of spraying. 

\subsection{The Survival of Original, Female Adults}

How many of the original flies survive spraying? If a fly survives one spray
cycle with probability $\phi$, then the probability that it survives $s$
consecutive sprays is $\phi^s$, assuming the probability of survival for each
spray is identical. The fly must also survive the normal hazards of life for the
$( s - 1 ) \sigma$ days from the first through to the last spray. The maximum
number of females from the original population which survive to the conclusion
of spraying, is therefore
\begin{eqnarray} \label{3}
N \ e^{ - \delta_2 ( s - 1 ) \sigma } \phi^{s}, 
\end{eqnarray} 
in which $N$ is the original, steady-state, equilibrium number of females prior
to spraying and $\delta_2$ is the worst-case-scenario, adult mortality rate. 

A simplification made in this formula is that no age distribution profile has
been assumed for the natural mortality. It should, however, be pointed out that,
while the original proportion of females which have not yet ovulated is
significant, the duration of the time preceding ovulation is insignificant when
compared to the length of the spraying operation itself. The average female fly,
yet to ovulate, will also already be of an age greater than zero and the full
time from eclosion to ovulation is therefore not under consideration, rather
some fraction of it. A further fact to bear in mind is that, in the field,
tenerals are not attracted in the same proportions as adult flies when measuring
the size of the original, equilibrium population. Of course, in the final
analysis, the natural mortality used matters little as the spray-survival rate,
$\phi$, is a small number. The chances of any of the original flies surviving
several spray cycles are usually practically zero. 

\subsection{The Survival of the Female Flies Which Eclosed from Original Pupae}

How many flies initially eclosed from such pupae? Assuming a population which was in equilibrium at some mean temperature prior to the commencement of spraying, the daily
number of flies ecloding from pre-spray-deposited pupae, is a constant
\begin{eqnarray} 
E_{\mbox{\scriptsize pre-spray}} &=& \beta N, \nonumber 
\end{eqnarray} 
in which $\beta$ is the steady-state eclosion rate previously described. Such flies continue to eclode for a period of one puparial duration subsequent to the commencement of spraying.

How many spray cycles will a given cohort be subjected to? The total number of spray cycles that a fly will be subjected to is determined by its day of eclosion, $\breve t$. The time, during the operation, that it spent above ground is the length of the operation less the time before eclosion, that is $\sigma ( s - 1 ) - \breve t$. The total number of insecticidal spraying cycles the fly will be subjected to is one more than the number of times a complete spray cycle fits into the period spent above ground. More succinctly,
\begin{eqnarray} 
\mathop{\rm floor} \left\{ \frac{\sigma ( s - 1 ) - \breve{t}}{\sigma} \right\} + 1, \nonumber
\end{eqnarray} 
where $\mathop{\rm floor}\left\{ . \right\}$ is the greatest integer function and $\breve{t}$ is the time from the first spray cycle to eclosion. The spray-survival rate, $\phi$, must be applied this many times, so that the fraction of flies which survives the entire operation is
\begin{eqnarray} 
\phi^{\mathop{\rm floor} \left\{ \frac{\sigma ( s - 1 ) - \breve{t}}{\sigma} \right\} + 1}. \nonumber
\end{eqnarray} 
What of natural mortality? The flies die off naturally at some age-dependent mortality, \\ \mbox{$\delta^*(t - \breve{t}, T)$}. 

What is the total number of flies of such origins remaining at the end of spraying? Collecting the above three observations, the number of female flies surviving at some later time, $t$, is 
\begin{eqnarray} 
\gamma \ \sum_{\breve{t} = 1 }^{\tau_0(T)} E_{\mbox{\scriptsize pre-spray}} e^{- \delta^* (t - \breve{t}, T)} \phi^{\mathop{\rm floor} \left\{ \frac{\sigma ( s - 1 ) - \breve{t}}{\sigma} \right\} + 1 }. \nonumber 
\end{eqnarray} 
At the completion of the operation, the total time elapsed is $( s - 1 ) \sigma$, and taking cognisance of the fact that the number of flies ecloding from pre-spray-deposited pupae must be constant for a population which was in equilibrium at some mean temperature, prior to the commencement of spraying, yields
\begin{eqnarray} \label{120}
\gamma \beta N \sum_{\breve{t} = 1 }^{\tau_0(T)} e^{- \delta^* (( s - 1 ) \sigma - \breve{t}, T)} \phi^{\mathop{\rm floor} \left\{ \frac{\sigma ( s - 1 ) - \breve{t}}{\sigma} \right\} + 1 }. 
\end{eqnarray} 
What if the length of the operation is less than one puparial duration? In the
event that the operation is curtailed to such an extent that the cycles are
terminated before $\tau_0$, then not all the original pupae have the opportunity
to eclode as flies and the remaining fraction contribute to the pupal
population, still in the ground at the end of spraying. Under these
circumstances, the above summation is truncated so that the upper limit,
$\tau_0(T)$ is replaced with $\sigma ( s - 1 )$. A further, extraordinary, pupal
contribution must then also be added to the tally of pupae, still present in the
ground at the end of spraying.

\subsection{The Production and Survival of Female Flies from inter-spray Pupae}

The last of the category ``original pupae'' eclode the moment one puparial duration since the commencement of spraying has
elapsed. All the flies
ecloding thereafter are of an inter-spray-larviposited origin. If, however,
spraying is curtailed to the extent that its duration is less than one puparial
duration, then none of this latter category ever eclode. Under such
circumstances they exist solely as pupae, still in the ground at the end of
spraying. 

Otherwise, the survival of flies ecloding from inter-spray pupae can be deduced by similar reasoning to the aforementioned case, one difference being that the number of emergent flies is
no longer constant over time (the ecloding population no longer being in equilibrium, or constant). Contributions to the time-$\breve t$-ecloding cohort arise as a result of pupae which
were larviposited $\breve{t} - \tau_0$ days before. The number of such flies at the conclusion of spraying is
\begin{eqnarray}\label{122}
\gamma \sum_{\breve{t} = \tau_0(T) + 1}^{\sigma ( s - 1 )} E(\breve{t}) \ e^{-\delta^*( ( s - 1 )\sigma - \breve{t}, T)} \phi^{\mathop{\rm floor} \left\{ \frac{\sigma ( s - 1 ) - \breve{t}}{\sigma} \right\} + 1 }. 
\end{eqnarray} 
The pupae were deposited by the previous, two survival categories and the inter-spray pupae, themselves. That is,
\begin{eqnarray}
E(\breve{t}) &=& E_a(\breve{t}) + E_{ps}(\breve{t}) + E_{is}(\breve{t}), \nonumber 
\end{eqnarray} 
in which the time-$\breve{t}$-ecloding cohorts are defined in Table
\ref{symbols}, according to their ancestral origins. 

What of the `knock-down' approach to the aerial spraying of tsetse? For
instances in which the duration of an operation has been curtailed to the length
of one puparial duration, or less, there is clearly no such contribution to
flies, only pupae. Under such circumstances this second and higher generation
category of flies may be completely disregarded. They need only be considered
from the point of view of a pupal population.

\subsubsection{The $E_a(\breve{t})$ Contribution to $E(\breve{t})$}

This is the contribution attributed to larviposition by original, adult females,
those which existed as such prior to the commencement of spraying and which
larviposit during the operation. By far the largest mass of the pupae
larviposited by original adults are larviposited during the first spray cycle,
between the first and second sprays. Their eclosion commences immediately after
the last of the pre-spray-larviposited pupae have emerged. They and a varying
proportion of the pupae larviposited during the second cycle, eclode during the
aerial spraying, for a completed operation. The majority of them are exposed to
the last, or last two, sprays, for such a completed operation. Terminating the
operation one spray short allows all the pupae larviposited in the second cycle
and a varying proportion of those larviposited during the first cycle, never to
be sprayed, in theory.

How many original mothers larviposit on a given day during the spraying? If
there were $N$ females prior to spraying, to assume that all have already
ovulated is a wost-case scenario, therefore a safe assumption. Inseminated
females, all $\eta N$ of them, are expected to deposit one pupa every $\tau_2$
days; that is, the larviposition of $\frac{\eta N }{ \tau_2 }$ pupae every day. 
\begin{figure}[H]
    \begin{center}
\includegraphics[height=5.3cm, angle=0, clip = true]{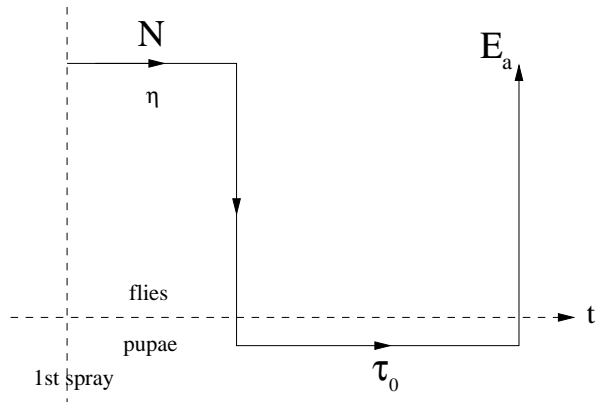}
\caption{Schematic diagram of second generation flies eclosed from pupae that were larviposited during spraying by original, pre-spray-existing adults.} \label{interSprayFromSurvivingAdults}  
   \end{center}
\end{figure}
How many of these potential mothers survive until a given day into the spraying operation? The proportion of these adult mothers which survive
naturally as long as $\breve{t} - \tau_0$ into spraying is $e^{- \delta_2 (
\breve{t} - \tau_0 )}$ and they are, in turn, subjected to 
\begin{eqnarray}
\mathop{\rm floor} \left\{ \frac{\breve{t} - \tau_0 - 1 }{\sigma} \right\} + 1 \nonumber
\end{eqnarray}
sprays (by contemplating Figure \ref{interSprayFromSurvivingAdults} and
assuming larviposition is successfully accomplished shortly before spraying on
the day in question). If a fly survives one spraying cycle with probability
$\phi$, then the probability that it survives the above number of cycles is 
\begin{eqnarray} 
\phi^{ \mathop{\rm floor} \left\{ \frac{\breve{t} - \tau_0 - 1 }{\sigma} \right\} + 1 }, \nonumber
\end{eqnarray} 
always assuming the probability of survival for each cycle is identical. 

How many of their daughters, in turn, survive to eclode? Taking natural mortality into account, the proportion of their pupae which survive to eclode is $e^{- \delta_0 \tau_0}$. 

Hence, the final expression
\begin{eqnarray} \label{103}
E_a(\breve{t}) &=& \eta \frac{ N }{ \tau_2 } \ e^{- \delta_2 ( \breve{t} - \tau_0 ) - \delta_0 \tau_0} \phi^{\mathop{\rm floor} \left\{ \frac{\breve{t} - \tau_0 - 1}{\sigma} \right\} + 1} \ H( \breve{t} - \tau_0 ),
\end{eqnarray}
in which $H$ is the version of the Heaviside step function with $H(0) = 0$. One prerequisite for such $E_a$ contributions to a second generation of flies is a restriction on the eclosion of the cohorts, $\breve{t} > \tau_0$ (again, by contemplating Figure
\ref{interSprayFromSurvivingAdults}). Otherwise they need only be considered from the point of view of a pupal population.
 
\subsubsection{The $E_{ps}(\breve{t})$ Contribution to $E(\breve{t})$} \label{Eps}

This is the contribution attributed to larviposition by mothers which existed as
pupae at the commencement of spraying. Many such pupae eclode subsequent to the
last spray, even for a completed operation. Under normal circumstances, this
category may be thought of as the problem category. How many such mothers come
into existence on a given day during the operation? The number of potential
mothers, ecloding daily (for a limited period), from pre-spray-deposited pupae
that will subsequently be inseminated, is
\begin{eqnarray} 
\gamma \eta E_{\mbox{\scriptsize pre-spray}} = \gamma \eta \beta N, \nonumber 
\end{eqnarray} 
in which $\beta$ is the steady-state, maximum possible, eclosion rate previously described, $N$ is the original, steady-state, equilibrium number of females prior to spraying, $\gamma$ is
the probability of being female and $\eta$ is the probability of insemination. Mothers of this category cease ecloding one puparial duration into the operation.

\begin{figure}[H]
   \begin{center}
\includegraphics[height=5.3cm, angle=0, clip = true]{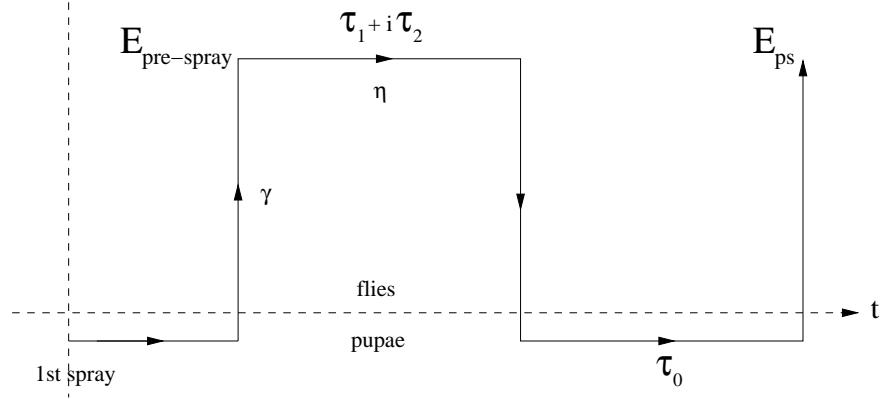}
\caption{Schematic diagram of second generation flies ecloding from pupae that were larviposited by flies from original pupae that existed as such at the commencement of spraying.} \label{interSprayFromPreSpray}
   \end{center}
\end{figure}

What is the subsequent mortality of these mothers? These pre-spray-larviposited mothers suffer a daily natural mortality of $\delta^*( \tau_1 + i \tau_2, T )$ and, by contemplating Figure
\ref{interSprayFromPreSpray}, are subjected to a total of 
\begin{eqnarray} \mathop{\rm floor}
\left\{ \frac{ \breve{t} - \tau_0 - 1 }{\sigma} \right\} - \mathop{\rm floor} \left\{
\frac{ \breve{t} - \tau_0 - \tau_1 - i \tau_2 - 1 }{\sigma} \right\} \nonumber
\end{eqnarray}   
sprays, this being the difference between the total number of sprays to
larviposition and the total number of sprays up to the day before the mother's
eclosion. If a fly survives one spraying cycle with probability $\phi$, then the
probability that it survives the above number of cycles is 
\begin{eqnarray} 
\phi^{ \mathop{\rm floor} \left\{ \frac{ \breve{t} - \tau_0 - 1 }{\sigma} \right\} - \mathop{\rm floor} \left\{ \frac{ \breve{t} - \tau_0 - \tau_1 - i \tau_2 - 1 }{\sigma} \right\} }, \nonumber
\end{eqnarray}
always assuming the probability of survival for each cycle is identical and that larviposition will be successfully accomplished before spraying on relevant days. The survival of such mothers is therefore readilly quantifiable in terms of the above. 

What are the temporal restrictions on the eclosion of the second generation
cohorts these mothers produce? By contemplating Figure
\ref{interSprayFromPreSpray}, the first requirement for second-generation
descent from such mothers, is a restriction on the cohorts to \mbox{$\breve{t} >
\tau_0 + \tau_1$}. The mothers would otherwise have had to have eclosed prior to
the first spray, a fact which would exclude them from the category presently
under consideration, altogether. Secondly, only for a limited period of time
(one puparial duration) do mothers which originate from pre-spray-deposited
pupae continue to emerge from the ground. That is, if all $\tau$s are integer
cohorts,
\begin{eqnarray} 
1 \ \le \ \breve{t} - \tau_0 - \tau_1 - i \tau_2 &\le& \tau_0  \hspace{10mm} i = 0, \ 1, \ \ldots \ , \nonumber 
\end{eqnarray} 
yielding a restriction on $i$,
\begin{eqnarray}
i &\le& \mathop{\rm floor} \left\{ \frac{1}{\tau_2} ( \breve{t} - \tau_0 - \tau_1 - 1 ) \right\}, \nonumber 
\end{eqnarray} 
and completing those on the time of eclosion,
\begin{eqnarray}
\tau_0 + \tau_1 + i \tau_2 \ < \ \breve{t} \ \le \ 2 \tau_0 + \tau_1 + i \tau_2. \nonumber 
\end{eqnarray} 
Lastly, only an $e^{- \delta_0 \tau_0}$ fraction of the pupae survive to eclode. 
Collecting all of the above information
\begin{eqnarray} \label{101} 
E_{ps}(\breve{t}) &=& \gamma \eta \beta N \sum_{i = 0}^{\mathop{\rm floor}\left\{ \frac{1}{\tau_2} ( \breve{t} - \tau_0 - \tau_1 - 1 ) \right\} } \left[ e^{- \delta^*( \tau_1 + i \tau_2, T ) - \delta_0 \tau_0 } \frac{}{} \phi^{ \mathop{\rm floor} \left\{ \frac{ \breve{t} - \tau_0 - 1 }{\sigma} \right\} - \mathop{\rm floor} \left\{ \frac{ \breve{t} - \tau_0 - \tau_1 - i \tau_2 - 1 }{\sigma} \right\} } \right. \hspace{5mm} \nonumber \\ 
&& \hspace{36mm} \left. \frac{}{} \left[ 1 - H( \breve{t} - 2 \tau_0 - \tau_1 - i \tau_2 ) \right] \ H( \breve{t} - \tau_0 - \tau_1 - i \tau_2 ) \right], 
\end{eqnarray} 
in which $H$ is the version of the Heaviside step function with $H(0) = 0$.
Notice that the last Heaviside factor becomes a precaution once $i$ is greater
than zero, since it is derived from the same inequality used for the restriction
on $i$. Clearly there is no $E_{ps}$ contribution to flies for instances in
which the duration of the operation has been curtailed to, or below, the time
between parturition and the production of the first larva, although pupae of
this category will certainly exist.

\subsubsection*{Modifications for a Continuous Model}

What if a continuous rather than discrete model were to be entertained? What if
the $\tau$s had not been rounded off to integer cohorts? What if they involved
fractions of a day, instead? The $i$ would, nonetheless, still be integers in
such a model although, by analogous reasoning to that above,
\begin{eqnarray} 
0 \ < \ \breve{t} - \tau_0 - \tau_1 - i \tau_2 &<& \tau_0  \hspace{10mm} i = 0, \ 1, \ \ldots \ , \nonumber 
\end{eqnarray} 
This would lead to a modification of the upper bound in the above summation, one based on
\begin{eqnarray}
\max \{ i \} &<& \frac{1}{\tau_2} ( \breve{t} - \tau_0 - \tau_1 ), \nonumber \end{eqnarray} 
as well as the replacement of
\begin{eqnarray}
\left[ 1 - H( \breve{t} - 2 \tau_0 - \tau_1 - i \tau_2 ) \right] \hspace{10mm} \mbox{ with } \hspace{10mm} H( - \breve{t} + 2 \tau_0 + \tau_1 + i \tau_2 ). \nonumber
\end{eqnarray}
The new switch differs in that it turns off when the argument zero, instead of
immediately above it. So far as the number of sprays is concerned,
`the-moment-before' replaces the `the-day-before' of the discrete case, so
that the relevant factor becomes 
\begin{eqnarray} 
\phi^{ \mathop{\rm floor} \left\{ \frac{ \breve{t} - \tau_0 }{\sigma} \right\} - \mathop{\rm floor} \left\{ \frac{ \breve{t} - \tau_0 - \tau_1 - i \tau_2 }{\sigma} \right\} }. \nonumber
\end{eqnarray} 

\subsubsection{The $E_{is}(\breve{t})$ Contribution to $E(\breve{t})$} \label{Eis}

\begin{figure}[H]
    \begin{center}
\includegraphics[height=5cm, angle=0, clip = true]{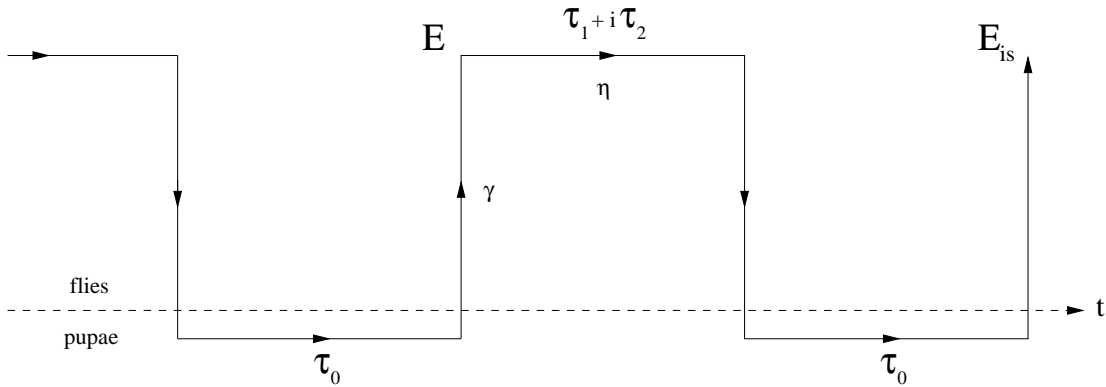} 
\caption{Schematic diagram of flies emerging from inter-spray pupae that are descended from inter-spray pupae themselves (third generation and higher).} \label{interSprayFromInterSpray} 
\end{center} 
\end{figure}

This is the contribution attributed to female flies descended from the mothers
which were themselves larviposited during spraying. They are the immediate
descendants of the $E_a$ category, the $E_{ps}$ category, or this very same
$E_{is}$ category itself. The first prerequisite for such third, or greater,
generation contributions is that $\breve{t} > 2 \tau_0 + \tau_1$ (by
contemplating Figure \ref{interSprayFromInterSpray}). 
Secondly, inter-spray-deposited pupal mothers only begin to eclode at time $\tau_0 + 1$, (just when the pre-spray pupae, have ceased to eclode). That is, if the various $\tau$s are integer cohorts, 
\begin{eqnarray} 
\tau_0 + 1 &\le& \breve{t} - \tau_0 - \tau_1 - i \tau_2 \hspace{10mm} i = 0, 1, ... \ , \nonumber
\end{eqnarray} 
yielding a restriction on $i$,
\begin{eqnarray} 
i &\le& \mathop{\rm floor} \left\{ \frac{1}{\tau_2} ( \breve{t} - 2 \tau_0 - \tau_1 - 1 ) \right\}. \nonumber
\end{eqnarray} 
The probability that such mothers survive the relevant number of spray cycles is
formulated in the same way as in the previous case; as is the natural mortality.
The pupal mortality of the mothers and the mortality of the grandmothers is
already taken care of by the $E_a$ and $E_{ps}$ categories. Taking cognizance of
the fact that the emergent population is not constant over time under such
circumstances, 
\begin{eqnarray} \label{102}
E_{is}(\breve{t}) &=& \gamma \eta \sum_{i = 0}^{\mathop{\rm floor}\left\{ \frac{1}{\tau_2} ( \breve{t} - 2 \tau_0 - \tau_1 - 1 ) \right\}} \left[ E( \breve{t} - \tau_0 - \tau_1 - i \tau_2 ) \ e^{- \delta^* ( \tau_1 + i \tau_2, T ) - \delta_0 \tau_0 } \hspace{20mm} \frac{}{} \right. \nonumber \\ 
&& \hspace{25mm} \left. \phi^{ \mathop{\rm floor} \left\{ \frac{ \breve{t} - \tau_0 - 1 }{\sigma} \right\} - \mathop{\rm floor} \left\{ \frac{ \breve{t} - \tau_0 - \tau_1 - i \tau_2 - 1 }{\sigma} \right\} } \ H( \breve{t} - 2 \tau_0 - \tau_1 - i \tau_2 ) \right]. 
\end{eqnarray} 
Notice, once again, that the Heaviside factor becomes a precaution once $i$ is
greater than zero, since it is derived from the same inequality used for the
restriction on $i$. There is no $E_{is}$ contribution to flies for instances in
which the duration of the operation is equal to, or below, the length of two
puparial durations and the time to the first larva. In fact, normal
circumstances make it difficult to imagine the category $E_{is}$ as ever having
eclosed by the end of spraying, therefore as having any relevance to the total
fly tally at all. $E_{is}$ may usually be neglected in the fly calculation.
Neither is there any $E_{is}$ contribution to pupae for instances in which the
duration of the spray operation is equal to, or below, the length of time
between parturition and the production of the first larva. $E$ may, in practice
and under normal circumstances, be assumed to have only two contributions,
$E_{a}$ and $E_{ps}$.

From this point on the origins of the inter-spray pupae are no longer relevant.
Generations higher than the third are accounted for through recursion, in
theory. In practice, the relative durations of the spraying operation, the
puparial stage and the time between eclosion and the production of the first
larva are such that it is difficult to imagine a scenario involving a fourth
generation, consequently any recurrence relation at all. 

\subsubsection*{Modifications for a Continuous Model}

What if a continuous rather than discrete model were to be entertained? What if
the $\tau$s had not been rounded off to integer cohorts? What if they involve
fractions of a day, instead? The $i$ would still be an integer in such a model,
however, by analogous reasoning to that above, 
\begin{eqnarray} 
\tau_0 &<& \breve{t} - \tau_0 - \tau_1 - i \tau_2 \hspace{10mm} i = 0, 1, ... \ , \nonumber
\end{eqnarray}
leading to a replacement of the upper bound in the above summation, one based on
\begin{eqnarray} 
\max \{ i \} &<& \frac{1}{\tau_2} ( \breve{t} - 2 \tau_0 - \tau_1 ). \nonumber
\end{eqnarray} 
So far as the number of sprays is concerned,
`the-moment-before' replaces the `the-day-before' of the discrete case, so
that the relevant factor becomes 
\begin{eqnarray}
\phi^{ \mathop{\rm floor} \left\{ \frac{ \breve{t} - \tau_0 }{\sigma} \right\} - \mathop{\rm floor} \left\{ \frac{ \breve{t} - \tau_0 - \tau_1 - i \tau_2}{\sigma} \right\} }, \nonumber
\end{eqnarray}
as in the previous case.

\section{Pupae Still in the Ground at the End of Spraying}

The total number of female pupae, which are still in the ground at the end of
spraying and {\bf which will survive to eclode}, is the $\gamma$ fraction of
flies destined to begin ecloding as a series of cohorts immediately subsequent
to the last spray. That is, starting at $\sigma (s - 1) + 1$, and ending with
$\sigma (s - 1) + \tau_0$, in the discrete case. Contributions to this pupal
population arise as a result of female pupae larviposited after the commencement
of spraying, in a completed operation. They may be categorized as: 
\begin{enumerate}
\item Daughters of original, female flies.
\item\label{daughtersOfPreSprayExistingPupae} Daughters of original, female pupae.
\item Third generation and higher daughters of females which were larviposited after the commencement of the spraying. 
\end{enumerate}
A fourth contribution, 
\begin{eqnarray} \label{320}
\gamma \beta N \sum_{\breve{t} = \sigma (s - 1) + 1}^{\tau_0(T)} e^{- \delta^* (( s - 1 ) \sigma - \breve{t}, T)} \phi^{\mathop{\rm floor} \left\{ \frac{\sigma ( s - 1 ) - \breve{t}}{\sigma} \right\} + 1 } H( \tau_0 - ( s - 1 ) \sigma ), \nonumber 
\end{eqnarray}  
that due to the pressence of original pupae, must also be taken into account in
an operation which has been curtailed to the extent that its duration is less
than one puparial duration. 

Otherwise, the total number of such female pupae remaining in the ground at the end of spraying and {\bf which will survive to eclode}, is 
\begin{eqnarray} \label{10} 
\gamma \sum_{\breve{t} = \sigma ( s - 1 ) + 1 }^{\sigma ( s - 1 ) + \tau_0(T)} \left[ E_a(\breve{t}) + E_{ps}(\breve{t}) + E_{is}(\breve{t}) \right].  
\end{eqnarray}  
As it transpires, one of the above categories is far and away more important
than any of the others in a completed operation. The pre-eminent category is
the second one above, the pupae which are destined to give rise to an $E_{ps}$
eclosion after spraying. The implications of this discovery are that, under
certain conditions, one formula can be adapted to provide a good estimate of the
outcome of aerial spraying. This fact is revealed when considering that there is
only one $O(\phi)$ contribution and this observation is further corroborated by
the algorithm of Childs\nocite{Childs4} (2011). A compositional analysis of the
origins of female pupae, still in the ground, reveals that summation of the
second term in the above summation formula is a good indicator of the entire
outcome of spraying, given a kill rate of 99.9\%, or better. It accounts for
well over 90\% of the pupal population at a kill rate of 99\%. 

\section{An Example of a Manual Calculation} \label{manualCalculation}

At $24 \ ^\circ\mathrm{C}$, four sprays which define three spray cycles, of
length 14 days each, are required. The aerial spraying scenario at this
temperature is slightly simplified and lends itself favourably to manual
calculation for two reasons. The first is that the time to the second last spray
is, for all practical purposes, exactly one puparial duration. All the pupae
deposited during the first spray cycle therefore eclode during the last spray
cycle and the pupae still in the ground at the end of the operation were
deposited during the second and third spray cycles. The second reason is that
the spray operation ends early from a metabolic point of view, meaning that a
third generation never exists during spraying, as is so often the case. For the
aforementioned reasons any problems with the formulae should be relatively easy
to detect.

\subsubsection*{Surviving Flies}

Although the number of spray cycles is relatively small, the number of original
adults which survive is still insignificant, it being of $O(\phi^4)$. Eq.
\ref{3} can accordingly be dismissed as negligeable. This is usually the case in
a completed operation. Those pre-spray-deposited pupae which eclode for the
duration of the second cycle must survive only the last two sprays, instead of
three, and \mbox{Eq. \ref{120}} therefore becomes
\begin{eqnarray} 
\gamma \beta N \sum_{\breve{t} = 15}^{28} e^{- \delta^* (42 - \breve{t}, 24)} \phi^{\mathop{\rm floor} \left\{ \frac{42 - \breve{t}}{14} \right\} + 1 } \ + \ O(\phi^3). \nonumber 
\end{eqnarray} 
The only categories left to consider are the second and third generations,
calculated according to Eq. \ref{122}. Both the $E_{is}$ and $E_{ps}$ terms can
be dismissed as irrelevant to the fly population, since the length of the
operation is shorter than the time from parturition to the production of the
first larva. Relevant second generation flies are therefore all descended from
the original, pre-spray-existing flies, those which survived the first spray.
This $E_a$ contribution is also a significant, $O(\phi^2)$ contribution, since
the pupae were deposited during the first spray cycle and eclode for the
duration of the last spray cycle. Eq. \ref{122} therefore becomes
\begin{eqnarray} 
\gamma \sum_{\breve{t} = 29}^{42} \eta \frac{ N }{ 10 } \ e^{ - 0.024 ( \breve{t} - 28 ) - 0.01 \cdot 28} \phi^{\mathop{\rm floor} \left\{ \frac{\breve{t} - 28 - 1}{14} \right\} + 1} \cdot 1 \cdot e^{-\delta^*( 42 - \breve{t}, 24)} \phi^{\mathop{\rm floor} \left\{ \frac{42 - \breve{t}}{14} \right\} + 1 } \ + \ O(\phi^3). \nonumber
\end{eqnarray} 
If one very crudely approximates $e^{- \delta^* (42 - \breve{t}, 24)}$ as $0.5231$, $e^{ - 0.024 ( \breve{t} - 28 ) - 0.01 \cdot 28}$ as $0.6389$ and $e^{-\delta^*( 42 - \breve{t}, 24)}$ as $0.7320$, in a `back-of-an-envelope' fashion and based on the relevant Childs\nocite{Childs4} (2011) mortalities, the sum of the preceding two expressions becomes
\begin{eqnarray} 
14 \times 0.5 \times N \left( 0.0466 \times 0.5231 \ + \ 1 \times \frac{ 1 }{ 10 } \times 0.6389 \times 0.7320 \right) \phi^2 \ + \ O(\phi^3), \nonumber
\end{eqnarray} 
using a $\beta$ of $0.0466$.

\subsubsection*{Surviving Pupae}

Only Eq. \ref{10} is relevant to the pupal outcome, since the operation is not
shorter than one puparial duration. The $E_{is}$ contribution can be dismissed
as irrelevant, since the length of the spray operation is shorter than $\tau_0 +
\tau_1$. In order for pupae to contribute to an $E_a$ eclosion subsequent to the
completion of the operation, they must have been larviposited in the second or
third cycles (those which were larviposited in the first cycle have already
eclosed by the end of spraying). This means that their mothers were sprayed at
least twice and accordingly they constitute an $O(\phi^2)$ contribution.
There is only one significant, $O(\phi)$ contribution to the pupal
population; that destined to give rise to an $E_{ps}$ eclosion.  The pupal
outcome can therefore be crudely formulated in terms of Eq. \ref{10} as
\begin{eqnarray} \label{6}
&& \hspace{-6mm} \eta \ \gamma^2 N \beta \sum_{{\breve t} = 43}^{70} \ \sum_{i = 0}^{ \mathop{\rm floor}\left\{ \frac{1}{10} ( \breve{t} - 28 - 16 - 1 ) \right\} } \left[ e^{- \delta( 28 + 16 + i 10, 24 ) } \phi^{ \mathop{\rm floor} \left\{ \frac{ \breve{t} - 28 - 1 }{14} \right\} - \mathop{\rm floor} \left\{ \frac{ \breve{t} - 28 - 16 - i 10 - 1 }{14} \right\} } \right. \hspace{10mm} \nonumber \\
&& \hspace{36mm} \left. \frac{}{} \left[ 1 - H( \breve{t} - 56 - 16 - i 10 ) \right] \ H( \breve{t} - 28 - 16 - i 10 ) \right]  \ + \ O(\phi^2). \nonumber
\end{eqnarray} 
This contribution arises as a result of mothers which eclode from original pupae
during the first and second cycles and which subsequently survive a single spray
to larviposit in the second and third cycles. Examination of the above formula
reveals significant, $O(\phi)$ terms only for the combinations of $i$ and
$\breve{t}$ represented in 
\begin{eqnarray} \label{6}
&& \hspace{-7mm} 1 \times 0.25 \times N \times 0.0466 \left[ \sum_{{\breve t} = 45}^{56} e^{- \delta( 44, 24 ) } \ + \ \sum_{{\breve t} = 59}^{70} e^{- \delta( 44, 24 ) } \right. \hspace{50mm} \nonumber \\
&& \hspace{50mm} \left. \ + \ \sum_{{\breve t} = 55}^{56} e^{- \delta( 44 + 10, 24 ) } \ + \ \sum_{{\breve t} = 69}^{70} e^{- \delta( 44 + 10, 24 ) } \right] \phi  \ + \ O(\phi^2). \nonumber
\end{eqnarray} 

\subsubsection*{Results}

Assuming the same mortalites and the same \mbox{8 000 000}, original,
steady-state number of females as in Childs\nocite{Childs4} (2011), the
estimated outcome for surviving, female pupae and flies is as presented in Table
\ref{handEstimatedResults}. 
\begin{table}[H]
\begin{center}\begin{tabular}{c|c c c c}  
&  &  &  & \\
$\phi$ & \ flies & $\mathop{log}($ flies $)$ & pupae & $\mathop{log}($ pupae $)$ \\ \\ \hline \\
0.01 \ & \ 398 & 2.60 & 12098 & 4.08 \\ \\
0.001 \ & \ 4 & 0.60 & 1210 & 3.08 \\ \\
0.0001 \ & \ 0 & - & 121 & 2.08 \\ \\
\end{tabular}
\caption{Estimated female survival for the simple case of $24 \
^\circ\mathrm{C}$, based on low order terms.} \label{handEstimatedResults}
\end{center}
\end{table}
The results only differ from those of the Childs\nocite{Childs4} (2011)
algorithm insofar as a more cautious choice of $\beta$ has been made.

\section{Conclusions}

Repeated spray cycles are scheduled at intervals two days short of the time
between eclosion and the production of the first larva, 
\begin{eqnarray} 
\sigma &=& \frac{ 1 }{ 0.061 + 0.0020 \left( T - 24 \right) } - 2, \nonumber
\end{eqnarray} 
and, in a completed operation, continue until two sprays subsequent to the
eclosion of the last, pre-spray-deposited, female pupae. That is,
\begin{eqnarray}
s &=& \mathop{\rm ceil} \left\{ \frac{1 + e^{5.5 - 0.25 T}}{0.057 \ \sigma} \right\} + 2, \nonumber 
\end{eqnarray}
in which $\mathop{\rm ceil}\left\{ . \right\}$ is the least integer function. 

Spray efficacy is found to come at a price due to the greater number of cycles
necessitated by cooler weather. The greater number of cycles is a consequence of
a larger ratio of puparial duration to time-to-first-larva at lower
temperatures. The prospect of a more expensive spraying operation at low
temperature, due to a greater, requisite number of spray cycles is, however, one
which is never confronted in the real world. In reality, one has to strive
towards kill rates and the only way such rates can be attained is by spraying at
as low a temperature as possible (Hargrove\nocite{Hargrove11}, 2009). 

A refinement of the existing formulae for the puparial duration and the time
between eclosion and the production of the first larva might be prudent in the
South African context of a sympatric {\em G. brevipalpis}-{\em G. austeni},
tsetse population. 

\subsection{The Complete Set of Formulae}

The complete set of formulae derived for the performance of a tsetse population under conditions of aerial spraying is summarised as follows.

\subsubsection*{Pupae}

The following are the contributions to female pupae, still in the ground at the end of spraying, {\bf which will survive to eclode}. The number of such pupae which are daughters of original adults is
\begin{eqnarray} \label{8}
\eta \gamma \frac{ N }{ \tau_2 } \sum_{{\breve t} = \sigma (s - 1) + 1}^{\sigma (s - 1) + \tau_0} \ e^{- \delta_2 ( \breve{t} - \tau_0 ) - \delta_0 \tau_0} \phi^{\mathop{\rm floor} \left\{ \frac{\breve{t} - \tau_0 - 1}{\sigma} \right\} + 1} \ H( \breve{t} - \tau_0 ).
\end{eqnarray}
The number of pupae which are daughters of original pupae is
\begin{eqnarray} \label{6}
&& \hspace{-6mm} \eta \ \gamma^2 N \beta \sum_{{\breve t} = \sigma (s - 1) + 1}^{\sigma (s - 1) + \tau_0} \ \sum_{i = 0}^{ \mathop{rm floor}\left\{ \frac{1}{\tau_2} ( \breve{t} - \tau_0 - \tau_1 - 1 ) \right\} } \left[ e^{- \delta( \tau_0 + \tau_1 + i \tau_2, T ) } \phi^{ \mathop{\rm floor} \left\{ \frac{ \breve{t} - \tau_0 - 1 }{\sigma} \right\} - \mathop{\rm floor} \left\{ \frac{ \breve{t} - \tau_0 - \tau_1 - i \tau_2 - 1 }{\sigma} \right\} } \right. \hspace{5mm} \nonumber \\
&& \hspace{48mm} \left. \frac{}{} \left[ 1 - H( \breve{t} - 2 \tau_0 - \tau_1 - i \tau_2 ) \right] \ H( \breve{t} - \tau_0 - \tau_1 - i \tau_2 ) \right].
\end{eqnarray} 
The number of pupae which are daughters of inter-spray pupae is
\begin{eqnarray} \label{9} 
&& \eta \gamma^2 \sum_{{\breve t} = \sigma (s - 1) + 1}^{\sigma (s - 1) + \tau_0} \ \sum_{i = 0}^{\mathop{\rm floor} \left\{ \frac{1}{\tau_2} ( \breve{t} - 2 \tau_0 - \tau_1 - 1 ) \right\}} \left[ E( \breve{t} - \tau_0 - \tau_1 - i \tau_2 ) \ e^{- \delta( \tau_0 + \tau_1 + i \tau_2, T ) } \hspace{20mm} \frac{}{} \right. \nonumber \\ 
&& \hspace{35mm} \left. \phi^{ \mathop{\rm floor} \left\{ \frac{ \breve{t} - \tau_0 - 1 }{\sigma} \right\} - \mathop{\rm floor} \left\{ \frac{ \breve{t} - \tau_0 - \tau_1 - i \tau_2 - 1 }{\sigma} \right\} } \ H( \breve{t} - 2 \tau_0 - \tau_1 - i \tau_2 ) \right]. 
\end{eqnarray} 
If the aerial spraying operation is curtailed to the extent that it is shorter than one puparial duration, then an additional category of pupae must be accounted for. The number of original pupae, those larviposited before the commencement of spraying, which have not yet eclosed by the end of spraying, is
\begin{eqnarray} \label{420}
\gamma \beta N \sum_{\breve{t} = \sigma (s - 1) + 1}^{\tau_0(T)} e^{- \delta^* (( s - 1 ) \sigma - \breve{t}, T)} \phi^{\mathop{\rm floor} \left\{ \frac{\sigma ( s - 1 ) - \breve{t}}{\sigma} \right\} + 1 } H( \tau_0 - ( s - 1 ) \sigma ). 
\end{eqnarray}

\subsubsection*{Flies}

The following are the contributions to female flies which survive to the
conclusion of spraying. The maximum number of such surviving, female flies from the original population, is 
\begin{eqnarray} \label{43}
N \ e^{ - \delta_2(T) ( s - 1 ) \sigma } \phi^{s}. 
\end{eqnarray} 
The number of surviving female flies which eclosed during spraying from original pupae is
\begin{eqnarray} \label{4}
\gamma \beta N \sum_{\breve{t} = 1 }^{ \min\{ \tau_0(T), \ \sigma ( s - 1 ) \} } e^{- \delta^* (( s - 1 ) \sigma - \breve{t}, T)} \phi^{\mathop{\rm floor} \left\{ \frac{\sigma ( s - 1 ) - \breve{t}}{\sigma} \right\} + 1 }. 
\end{eqnarray} 
The number of surviving female flies that eclosed from
inter-spray-larviposited pupae and which will survive until after the last
spray is
\begin{eqnarray} \label{5}
\gamma \sum_{\breve{t} = \tau_0(T) + 1}^{\sigma ( s - 1 )} \left[ E_a(\breve{t}) + E_{ps}(\breve{t}) + E_{is}(\breve{t}) \right]  \ e^{-\delta^*( ( s - 1 )\sigma - \breve{t}, T)} \phi^{\mathop{\rm floor} \left\{ \frac{\sigma ( s - 1 ) - \breve{t}}{\sigma} \right\} + 1 },
\end{eqnarray} 
in which the respective time-$\breve{t}$-ecloding cohorts are given by
\begin{eqnarray}
E_a(\breve{t}) &=& \eta \frac{ N }{ \tau_2 } \ e^{- \delta_2 ( \breve{t} - \tau_0 ) - \delta_0 \tau_0} \phi^{\mathop{\rm floor} \left\{ \frac{\breve{t} - \tau_0 - 1}{\sigma} \right\} + 1} \ H( \breve{t} - \tau_0 ), \nonumber \\
E_{ps}(\breve{t}) &=& \gamma \eta \beta N \sum_{i = 0}^{\mathop{\rm floor}\left\{ \frac{1}{\tau_2} ( \breve{t} - \tau_0 - \tau_1 - 1 ) \right\} }
\left[ e^{- \delta( \tau_0 + \tau_1 + i \tau_2, T ) } \frac{}{}
\phi^{ \mathop{\rm floor} \left\{ \frac{ \breve{t} - \tau_0 - 1 }{\sigma} \right\} - \mathop{\rm floor} \left\{ \frac{ \breve{t} -
\tau_0 - \tau_1 - i \tau_2 - 1 }{\sigma} \right\} } \right. \nonumber
\\ 
&& \hspace{38mm} \left. \frac{}{} \left[ 1 - H( \breve{t} - 2 \tau_0 - \tau_1 - i \tau_2 ) \right] \ H( \breve{t} - \tau_0 - \tau_1 - i \tau_2 ) \right], \nonumber \\ 
E_{is}(\breve{t}) &=& \gamma \eta \sum_{i = 0}^{\mathop{\rm floor}\left\{ \frac{1}{\tau_2} ( \breve{t} - 2 \tau_0 - \tau_1 - 1 ) \right\}} \left[ E( \breve{t} - \tau_0 - \tau_1 - i \tau_2 ) \ e^{- \delta( \tau_0 + \tau_1 + i \tau_2, T ) } \frac{}{} \right. \nonumber \\ 
&& \hspace{29mm} \left. \phi^{ \mathop{\rm floor} \left\{ \frac{ \breve{t} - \tau_0 - 1 }{\sigma} \right\} - \mathop{\rm floor} \left\{ \frac{ \breve{t} - \tau_0 - \tau_1 - i \tau_2 - 1 }{\sigma} \right\} } \ H( \breve{t} - 2 \tau_0 - \tau_1 - i \tau_2 ) \right]. \nonumber 
\end{eqnarray}

\subsubsection*{Modifications for a Continuous Model}

Minor modifications, listed at the ends of Subsections \ref{Eps} and \ref{Eis}, need to be made if the discrete formulae are to be adapted to the continuous case. The summations over the cohorts $\sum_{{\breve t} = 1}^{\ldots}$, $\sum_{\breve{t} = \tau_0(T) + 1}^{\ldots}$ and $\sum_{{\breve t} = \sigma (s - 1) + 1}^{\ldots}$ would also be replaced with the integrals $\int_{0}^{\ldots}$, $\int_{\tau_0(T)}^{\ldots}$ and $\int_{\sigma (s - 1)}^{\ldots}$, respectively and among other things.

\subsection{Which Formulae are Significant, Which are Insignificant and Which are Irrelevant?}

The pupae which give rise to the $E_{is}$ eclosion have, at best, hardly come
into existence, let alone eclosed, by the end of a completed operation. The
$E_{is}$ contribution may therefore usually be regarded as irrelevant insofar as
flies are concerned, whereas its contribution to the total pupal tally is
usually insignificant, at most. The category $E_{is}$ can be omitted from the
\mbox{Eq. \ref{5}} fly formula for all reasonable circumstances, that is, unless
the operation has been greatly extended. Clearly, there is no $E_{is}$
contribution to the Eq. \ref{5} fly formula for instances in which the duration
of the operation is $2 \tau_0 + \tau_1$, or less. There is also no Eq. \ref{9}
contribution involved in the pupal tally for instances in which the duration of
the operation is $\tau_0 + \tau_1$, or less. Although this varies from case to
case, the Eq. \ref{9} contribution is never significant unless the operation has
been greatly extended. The $E_{ps}$ term can be omitted from the Eq. \ref{5} fly
formula for instances in which the duration of the operation is $\tau_0 +
\tau_1$, or less. Take heed, however, that pupae, destined to give rise to a
future $E_{ps}$ eclosion (Eq. \ref{6}), are usually the most significant
contribution, by far, in a completed operation. The only circumstances for which
these pupae are irrelevant is for instances in which the duration of the
operation involves a single cycle; that is, two sprays only. The category
$E_{a}$ can be omitted from the Eq. \ref{5} fly formula for instances in which
the duration of the operation is $\tau_0$, or less. The Eq. \ref{8} pupae, those
destined to give rise to just such an $E_{a}$ eclosion, can never be irrelevant,
since the larviposition of such pupae commences immediately after the first
spray. A lengthly spray operation can, however, still render their contribution
insignificant, since such mothers become progressively decimated and the earlier
pupal mass will eclode before the operation is complete. Generally, the more
curtailed the spraying operation, the fewer contributing categories there are,
with one exception: In the event that an operation is curtailed to the extent
that its duration is less than one puparial duration, a proportion of the
original pupae remain in the ground at the end of spraying. This complicates
matters slightly. Under such circumstances the series of flies, Eq. \ref{120},
is truncated as specified in the formula Eq. \ref{4}, the remainder being
additional pupae which will survive, as quantified by Eq. \ref{420}. These same
circumstances render the entire Eq. \ref{5} fly formula irrelevant.

In a completed operation, by far the most significant category is that
calculated in terms of \mbox{Eq. \ref{6}} above. These are pupae which are
destined to give rise to an $E_{ps}$ eclosion once the operation is complete.
The magnitude of this contribution is easy to see when one considers that many
of their mothers (original pupae which eclode during spraying) will only be
sprayed once (see Fig. \ref{interSprayFromPreSpray}). Since all pupae are, by
definition, never sprayed if they are still in the ground at the end of
spraying, the $E_{ps}$ lineage constitutes an $O(\phi)$ contribution to pupae.
In contrast, the mothers of pupae destined to give rise to an $E_{a}$ eclosion,
after spraying, must be sprayed more than once if these daughters are still to
be pupae by the end of the operation. Otherwise, they would already have eclosed
(see Fig. \ref{interSprayFromSurvivingAdults}). The Eq. \ref{8} lineage may
therefore be regarded as being of $O(\phi^2)$ significance. The Eq. \ref{9}
lineage (pupae destined to give rise to an $E_{is}$ eclosion, after spraying)
also constitutes an $O(\phi^2)$ contribution (see Fig.
\ref{interSprayFromInterSpray}). A large number of the pupae destined to give
rise to an $E_{ps}$ eclsoion will therefore always be of a lower order than
those destined to give rise to an $E_{a}$ or an $E_{is}$ eclosion, in a
completed operation. Notice that all lineages which exist as flies at the time
of the last spray will be of order $O(\phi^2)$, or higher, in a completed
operation. This is easy to see when one considers that all these flies must, by
definition, be subjected to the last spray. The length of a complete operation
means that their lineage must also have been sprayed at least once during the
operation. That the daughters of the original pupae, Eq. \ref{6}, are a good
forecast of the outcome of a completed  operation, given a kill rate of 99.9\%
or better, is further corroborated by the algorithm of Childs\nocite{Childs4}
(2011).

Given the high kill rates attainable, it is not surprising that the outcome, for
flies (as distinct from pupae), is largely determined by the size of the
emergent population which was only subjected to the last two sprays, in a
completed operation. This is why the proportion of flies was relatively high in
the Section \ref{manualCalculation} example. A full cycle's worth of original
pupae eclosed to be followed by a full cycle's worth of $E_{a}$ eclosion during
the last two cycles of that example. The actual flies, themselves, which survive
the last spray of a completed operation are, however, of no real consequence to
the outcome. Under such circumstances, pupae, still in the ground at the end of
spraying, are identified as the main threat to successful control by aerial
spraying. The outcome, for kill rates of 99.9\%, or higher, was shown in
Childs\nocite{Childs4} (2011) to be almost exclusively dependent on the
immediate descendants of the original pupae, those pupae which were present at
the commencement of spraying. Even at kill rates as low as 99\% this Eq. \ref{6}
category still constitutes around 90\% of the surviving female population in a
completed operation while the Eq. \ref{8} contribution accounts for less than
10\% of the total pupal population (Childs\nocite{Childs4}, 2011). 

If, however, operations are halted one or more sprays short, these
generalisations can not be made. Not only is the recently-eclosed fly
population still significant, there is also a fairly large pupal population
descended from the original adults. The contribution of the Eq. \ref{8} category
and others, becomes significant. The total predominance of Eq. \ref{6} does not
exist. Some of the pupae destined to give rise to an $E_{a}$ eclosion become a
significant, $O(\phi)$ contribution, as do some of the flies that eclosed from
original pupae. While flies arising from the $E_{a}$ eclosion, itself, are only
of order $O(\phi^2)$, it should also be remembered that they were larviposited
during the first cycle, when the population of original, adult flies was still
strong. The formulae Eq. \ref{8}, Eq. \ref{6}, Eq. \ref{43}, Eq. \ref{4} and
part of Eq. \ref{5} all need to be considered if the operation is halted one
spray short. The Eq. \ref{6} contribution is always significant and that of Eq.
\ref{43} insignificant. Only the $E_{a}$ term in Eq. \ref{5} is still relevant
under such circumstances. If the last spray falls close to a full cycle's length
from the one-puparial-duration mark, then the Eq. \ref{4} flies will be less
significant, despite the fact that they are an $O(\phi)$ contribution, and one
will find almost a cycles worth of $O(\phi^2)$, $E_{a}$ eclosion (Eq. \ref{5}).
If, on the other hand, the last spray falls close to the one-puparial-duration
mark, then one will find almost a cycles worth of $O(\phi)$, Eq. \ref{4}
eclosion and the entire Eq. \ref{5} contribution can be dismissed as
insignificant. They will mostly still be in the ground as Eq. \ref{8} pupae. Of
course, a proportion of the alleged pupae might also actually be aging flies,
under such circumstances, given the mechanism whereby mature, gravid females
excrete lipophilic toxins to sacrifice larvae in utero for their own survival. 

If the operation is halted two, or more, sprays short, the Eq. \ref{5} fly
formula falls away entirely while an Eq. \ref{420} contribution comes into
existance. Under these circumstances, only the formulae Eq. \ref{8}, Eq.
\ref{6}, Eq. \ref{420}, Eq. \ref{43} and Eq. \ref{4} are relevant. The
significance of the Eq. \ref{420} contribution will be proportional to the time
between the last spray and the one-puparial-duration mark. Of course, the
original, surviving adults (Eq. \ref{43}) can almost always be regarded as
insignificant, they being an $O(\phi^s)$ contribution. That is, for all except
the most severely curtailed operation. 

\subsection{Factors Extraneous to a Theoretical Outcome}

It is important to remember that the formulae calculate a theoretical outcome
based on the premise that no practical problems will be encountered in the
field. The idea of this work has been to create a simple arithmetic tool which
can be used to establish conditions sufficient for a successful operation in the
context of a closed tsetse population. Hargrove\nocite{Hargrove10} (2005)
quantifies the dangers in allowing the smallest of founding populations to
survive and re-invasion is an ever present threat which will ultimately
compromise even the most successful aerial spraying operation. A cursory
inspection of the Rogers and Robinson \nocite{RogersAndRobinson} (2004) study
(based on the Ford and Katondo\nocite{FordAndKatondo}, 1977, maps) suggests that
most tsetse populations cannot be considered closed. The total extent of habitat
is a further cause for concern. Even the extant, forest-dwelling, tsetse
populations of South Africa cannot be considered closed and extend beyond its
borders (Hendrickx\nocite{Hendrickx}, 2007). By far the biggest threat to any
aerial spraying operation on mainland Africa is re-invasion from adjacent,
untreated areas. Closed populations need to be created by temporary barriers of
odour-baited targets such as the one used successfully by Kgori et
al.\nocite{Torr1}, 2006. Childs\nocite{Childs3} (2010) and Esterhuizen et
al.\nocite{EsterhuizenKappmeierGreenNevillVanDenBossche} (2006) comprehensively
researched the design of such odour-baited, target barriers for {\em G.
austeni} and {\em G. brevipalpis}; albeit mostly from a point of view of a
control in its own right. In Childs\nocite{Childs4} (2011), the same model was
re-run with a more stringent, {\em G. austeni} isolation standard than that
used for control in Childs\nocite{Childs3}, 2010.

Quantifying spray efficacy represents a further problem. Temperature doesn't
only effect the aerial spraying of tsetse, through its reproductive cycle and
general population dynamics. Cooler weather is preferred for aerial spraying
from a point of view of spray efficacy (\mbox{Hargrove\nocite{Hargrove11}},
2009). Very high kill rates usually (though not always) come about as a result
of the sinking air associated with cooler weather. It favours the settling of
insecticidal droplets. Although \mbox{Du Toit}\nocite{DuToit} (1954) makes
mention of the sustained down draught from a slow-moving helicopter, there are
obviously distinct disadvantages to such a method of insecticide application.
The inherent toxicity of deltamethrin and many other pyrethroids also decreases
with temperature, contrary to the toxicity of most insecticides. The effects of
temperature on spray efficacy are not modelled. For that matter, neither are the
effects of anabatic winds, nor the protection afforded by the forest canopy and
multifarious other variables relevant to spray efficacy. Spray efficacy is
usually measured in the field, with hindsight, rather than predicted. 

It has been assumed that gravid female resistance to the insecticide can be
ignored. Although it may be tempting to consider the model poorer for this lack
of detail, it may not be a defficiency of any consequence, since a simple
trade-off exists between a fly living and a larva dying. Further pregnancies
during the operation should, similarly, terminate in spontaneous abortion.
Another point to bear in mind is that the spray-survival rate is a small
fraction. Whatever the exact value of $\phi$ may be for the older flies, the
value of $\phi^s$, or similar, should ordinarily ensure that they have been
decimated. Gravid female resistance is sure to result in the use of slightly
altered eclosion rates and inappropriate natural mortalities, however, some
comfort can be taken from the knowledge that the effects of natural mortalities
are very small in comparison to those due to aerial spraying. They have little
bearing on the outcome.

\section{Acknowledgements} 

The original conception of this project was entirely John Hargrove's and it was
under his direction that the work originally commenced. The author is indebted
to Johan Meyer and Natalie Le Roux of the University of the Free State for
hosting this work. Lastly, both the editor and the anonymous reviewers are
thanked for their suggestions and guidance. 





\bibliography{aerialSprayingFormulaePaper}



\end{document}